\documentclass{iau}
\usepackage{natbib}

\newcommand{\aap}{    {\it Astron. Astrophys.}}

\newcommand{\apj}{    {\it Astrophys. J.}}
\newcommand{\apjl}{   {\it Astrophys. J. Lett.}}

\newcommand{\jgr}{    {\it Journal of Geophysical Research}}

\newcommand{\solphys}{{\it Solar Phys.}}

\title[Initiating and driving prominence eruptions] 
{The physical mechanisms that\\initiate and drive solar eruptions}

\author[Guillaume Aulanier]   
{Guillaume Aulanier}

\affiliation{
Observatoire de Paris, LESIA, CNRS, UPMC, Univ. Paris Diderot,\\
5 place Jules Janssen, 92190 Meudon, France
\\email: {\tt guillaume.aulanier@obspm.fr}
}

\pubyear{2013}
\volume{300}  
\pagerange{119--126}
\jname{Nature of Prominences and their Role in Space Weather}
\editors{A.C. Editor, B.D. Editor \& C.E. Editor, eds.}
\begin{document}

\maketitle

\begin{abstract}
Solar eruptions are due to a sudden destabilization of force-free coronal 
magnetic fields. But the detailed mechanisms which can bring the corona 
towards an eruptive stage, then trigger and drive the eruption, and 
finally make it explosive, are not fully understood. A large variety of 
storage-and-release models have been developed and opposed to each other since 
40 years. For example, photospheric flux emergence vs. flux cancellation, 
localized coronal reconnection vs. large-scale ideal instabilities and loss of 
equilibria, tether-cutting vs. breakout reconnection, and so on. The competition 
between all these approaches has 
led to a tremendous drive in developing and testing all these concepts, by 
coupling state-of-the-art models and observations. Thanks to these developments, 
it now becomes possible to compare all these models with one another, and to 
revisit their interpretation in light of their common and their different behaviors. 
This approach leads me to argue that no more than two distinct physical mechanisms 
can actually initiate and drive prominence eruptions: the {\em magnetic breakout} and 
the {\em torus instability}. In this view, all other processes (including flux 
emergence, flux cancellation, flare reconnection and long-range couplings) should 
be considered as various ways that lead to, or that strengthen, one of the 
aforementioned driving mechanisms. 
\end{abstract}

\firstsection 
\section{Introduction}

Eruptive prominences are large clouds of magnetized plasma, which are ejected 
from the low solar corona into interplanetary space, in the form of Coronal 
Mass Ejections (CMEs). They can erupt either from within active regions, or 
from long filament channels. During the eruption, the system accelerates up 
to typical velocities of $100-1000$ km/s (although slower and faster CMEs 
also exist) while flare loops always form in the wake of the eruption (even 
though they can be hard to see in weak events). 

Since the low corona is a sufficiently collisional plasma, its evolution can be 
studied in the frame of MHD.
Also, the ratio between thermal and magnetic pressure is there very small, i.e. $\beta\ll1$. 
Therefore, the magnetic energy 
dominates all other forms of energy in the source regions of solar eruptions 
\citep[see][Table 2]{Forbes00}.
Current-free (potential) magnetic fields correspond to the minimum 
magnetic energy for a given distribution of magnetic flux through the dense photosphere. 
Since the photospheric flux distribution does not significantly change during the
time-scales of eruptions, and since the powering of eruptions requires the magnetic 
energy to decrease, the coronal magnetic field must therefore be highly non-potential 
prior to eruption onset, i.e. it must contain strong electric currents.
Due to the slow evolution of the photospheric magnetic field (as compared to typical 
coronal velocities), currents which are injected into the corona must accumulate slowly, 
such that the coronal field evolves quasi-statically, as a sequence of force-free equilibria.
The triggering of CMEs therefore requires the coronal field to reach some threshold 
above which the balance between magnetic pressure (which points upward) and magnetic tension 
(which points downward) is broken. When the system suddenly enters a regime in which the 
pressure dominates, it can erupt 
in a catastrophic way, leading to a CME. The resulting ideal expansion of the magnetic 
field, as well as the resistively driven magnetic reconnection 
in the current layer that forms in the wake of the expanding system, both contribute 
to decrease the magnetic energy. These arguments are the root of the ``storage-and-release'' 
MHD models for solar eruptions. 

Even though it is now widely accepted that solar eruptions are due to such 
a violent destabilization of previously energized coronal magnetic fields, 
the detailed mechanisms which bring a system into an eruptive stage, and 
which eventually drive the eruption, are not yet fully understood. A large 
variety of storage-and-release models has been put forward in the past decades 
\citep[see][for two extensive reviews that also describe observations]
{Forbes06,Schmieder13}. Firstly, most of these models nicely describe many observed aspects of 
solar eruptions. Therefore it is difficult to estimate their respective merits solely 
based on observational criteria. Secondly, the models qualitatively share many 
common physical ingredients. So they may be difficult to distinguish from one 
another. So, to date, two questions remain open: Which 
{\em physical mechanisms} drive prominence eruptions? Which {\em solar drivers} 
can gradually bring stable prominences to eruptive states? 

This paper aims at reviewing the existing storage-and-release models that are realistic 
enough, in terms of the solar physical conditions, and at considering them all together 
in a common frame, so as to bring some up-to-date answers to the two aforementioned 
questions. So this review focuses only on the {\em onset and driving mechanisms} 
of eruptions, not on their ensuing development in the large-scale corona and in the 
heliosphere. 

\section{Non-equilibrium and instability electric-wire models}

The oldest prominence eruption model that remains 
considered to date is the loss-of-equilibrium model, that was initially 
put forward in the physical paradigm of electric-wires, and that was further 
proven to occur in fully 3D MHD simulations. 

\subsection{Straight-wire geometry}

The original model 
was developed in 2D, in cartesian geometry \citep{vantend78,vantend79,Moloden87,Filippov01}. 
The set-up consists of 
a line current $I$ that is inserted at some height $z=h$ above the photospheric
plane, $z=0$, an ambient coronal field $B_{\mathrm{ex}}$, and a so-called 
``image current'' $-I$ is added at $z=-h$
to emulate one effect
of photospheric line tying, i.e. so that 
the photospheric magnetic field does not change when $h$ changes. 
The resulting coronal magnetic field consists of a detached plasmoid (or flux rope) 
that mimics a prominence that is embedded in a coronal arcade, and whose apex is 
located at $z=h$. 

In the ``electric 
paradigm'', the equilibrium of the system 
results from the competition between two Laplace forces, namely the downward 
force that $B_{\mathrm{ex}}$ excerts on the coronal line current, and the upward 
force generated by the repulsion of the two line currents. In the ``MHD paradigm'', 
the former corresponds to restraining magnetic tension of the potential 
field overlying the flux rope, and the latter to magnetic pressure 
that results from the increase of 
the magnetic field strength below the coronal line current 
induced by the photospheric boundary. 

With these settings, the equilibrium curve $h(I)$ has a critical point $(I_c;h_c)$, 
beyond which the line-current $I \ge I_c$ cannot stay in equilibrium and 
must move to infinite $z$. 
The altitude $z=h_c$ of this critical point 
is given by the height at which 
$B_{\mathrm{ex}}(z)$ starts to drop 
faster than $z^{-1}$. 

The cartesian model has been refined several times, e.g. by giving a finite width to 
the coronal current, by taking into account the conservation of magnetic flux during 
the eruption, and by treatig the 
line-tying at the photospheric part of the flux rope. The latter yields 
the formation of a vertical current sheet below the flux rope during its eruption 
\citep{Mar89,AmariAly90}. 
This current sheet 
exerts an extra restraining force on the line current, such that
the flux rope cannot move to infinity, 
but finds a new equilibrium position at finite $z$ \citep{ForbesIsen91}.
In 2D, a full eruption requires the dissipation of this 
current sheet by sufficiently fast 
magnetic reconnection \citep{LinForbes00}. But this may not be required 
in 3D. This is in line with analytical MHD considerations 
on the energy of fully open (so unreconnected) magnetic fields. Indeed this 
energy is infinite in 2D cartesian geometry, while it remains finite in 2D 
axisymmetric spherical systems and in all 3D geometries \citep{AlyJJ84,AlyJJ91,Stu91}. 


\subsection{Curved-wire geometry}

The model has also been investigated in 2.5D axisymmetric (toroidal) geometry. 
In a first approach, the coronal line-current is replaced by a detached ring-current 
at some height above the photospheric spherical surface \citep{Lin98}. If an image current 
is added below the photosphere so that the coronal arcades surrounding 
the flux rope are line-tied, the same repulsive and restraining forces as 
discussed above contribute to the force balance. However, a new repulsive force 
(which the current exerts on itself due to its bending) comes into play. This 
curvature (or ``hoop'') force is radially outward directed and can be balanced 
by an external magnetic field, $B_{\mathrm{ex}}$ \citep{Shafra66,Chen89,TD99}. 

In these spherical models, the requirement for magnetic reconnection below the 
rope as identified in cartesian geometry \citep{LinForbes00} still holds, but it 
is less important because the rope can rise ideally to tens of solar radii before 
the Laplace force of the vertical current sheet can halt the eruption. 

In a second approach, half of the ring-current of radius $R$ is emerged above 
a planar photosphere, and the other half located below the photosphere somehow 
plays the role of the image current. With these settings, the untied ring-current 
can freely expand radially, as a result of a so-called ``torus instability''. 
This instability occurs when the restoring force due to the external field 
drops faster with the altitude than the hoop force. For external poloidal fields 
(i.e. perpendicular to the current) with $B_{\mathrm{ex}} \sim R^{-n}$, the instability 
threshold is given by $n_c \sim 3/2$ \citep{Bateman78,KliTor06}.

Qualitatively similar instability thresholds have been identified when the 
line-tying of the ring-current is treated, through the addition of multiple 
image current segments in the model \citep{IsenForbes07,Olmedo13b}. 

\subsection{Discussion on electric-wire models}

The cartesian and the axisymmetric models had initially been developed 
separately. The former studied the conditions for ``loss-of-equilibria'', 
and the latter calculated onset criteria for ``instabilities''. Both 
approaches were recently revisited by \citet{DemAula10}. Non-circular 
current paths were later considered \citep{Olmedo10,Olmedo13a,Olmedo13b}. 
All geometries were shown to share almost the same analytical equations, and 
therefore the same physics. It was then proposed to join both approaches in 
a single ``torus instability'' mechanism. 

This model has been criticized by several MHD physicists. 
Indeed the physical simplifcations of the electric-wire paradigm, and 
the qualitative nature of their link with the (correct) MHD paradigm, 
are a priori quite disputable. Nevertheless, the analytical elecric-wire 
predictions for eruptive thesholds have been found to match the onset 
of eruptions in some line-tied MHD simulations. Those include suspended flux 
ropes in 2.5D \citep{Forbes90} and in 3D \citep{Inoue06,Nishida13}, and fully 
3D line-tied flux ropes \citep{Roussev03,To05,TorKli07,SchrijverE08,Tor10,Jiang13}. 
So the electric-wire model was found to be consistent with its correct 
MHD treatment. But even then, some questions were left open. 
%
Indeed, even if all the aforementioned simulations correctly prescribed 
force-free flux ropes as initial conditions, 
firstly all but one used analytical flux rope solutions from \citet{TD99}
that contain very idealized current distributions 
(much simpler than those produced by solar MHD processes [see e.g. \citealt{Aula05a}], 
which are themselves more compatible with photospheric observations [see e.g. 
\citealt{SchmiederAula12,Georg12}]), and secondly these ropes were already unstable, 
so that their pre-eruptive evolution was not self-consistently treated. 

In spite of all these issues, the torus instability was found to occur in some 
recent MHD simulations in which 3D flux ropes were gradually formed by photospheric 
drivers that mimic solar processes \citep[see][as described further below]{Aula10,Fan10}. 
So the torus instability appears as a robust process to initiate and drive solar eruptions.

\section{MHD models based on increasing manetic pressure}

Any realistic eruption model must involve non-potential pre-eruptive coronal 
fields. There are various ways to generate them, as listed below. Some models 
investigated the role of the increasing magnetic pressure alone to drive an 
eruption. 

\subsection{Axial flux increase}

The first models that were developed in the correct MHD paradigm were 
analytical and two-dimensional. There the prominence axis was oriented 
perpendicularly to the 2D plane of the models, and the magnetic shear was 
substituted for the electric current $I$ as a primary variable. But in 
the absence of a self-consistent way to prescribe increasing magnetic 
shear along the prominence, these models rather prescribed the prominence 
axial magnetic field (or flux) as a free parameter. The stability properties 
of the modeled systems were analyzed, in pretty much the same way as in the 
electric-wire models. Equilibrium curves were identified, and the lack of 
existing solutions were found for specific parameters, in particular for 
strong axial fields and when thermal pressure was taken into account 
\citep{Low77,Birn78,Heyvaerts82,Zwing87}. 

These models remained theoretical, until the development of the 
flux-insertion method through magneto-frictional numerical relaxations 
in 3D \citep{van04}. This novel approach allowed to 
model observed prominences and to find some eruptions, by ``manually'' 
inserting axial fluxes of different prescribed magnitudes 
\citep{Su11}. 

The early 2D models and the recent 3D ones qualitatively interpreted their 
modeled eruptions as evidences for losses of equilibria that could occur 
when the ratio $\mathfrak{R}$ of the axial prominence flux to the overlying arcade 
flux exceeds some unidentified threshold \citep[as discussed by][]{Heyvaerts82,Green11}. 

It is only very recently that \citet{Kliem13} performed new detailed analysis 
of the 3D models. They found that the eruption onset condition matches the 
threshold as predicted by the electric-wire models, namely the torus instability. 
This result is important in two ways. Fistly it shows that, even though the ratio 
$\mathfrak{R}$ is defined from the right MHD paradigm, its unclear condition 
for eruptiveness has to be substituted by the clearer criterion for torus 
instability, even if that one comes from the disputable electric-wire paradigm. 
Secondly, this result provides one more case of torus instability in numerical 
simulations. 

\subsection{Line-tied shearing and twisting motions}

The development of 3D line-tied MHD simulations showed that, when the 
system is driven by horizontal photosphetic motions, the axial flux cannot 
increase arbitrarily. Two situations were identified. 

Firstly, if shearing or twisting flux tubes are restrained by strong 
non-moving overlying arcades in 3D, the axial flux eventually saturates. 
Then the system can either remain stable \citep{Ant94,Devore00,Aula02} 
or eventually develop a kink instability that subsequently disrupts the 
whole configuration \citep{Amari99}. 
Secondly, if the overlying arcades are either 
too weak or also sheared or twisted, the whole system starts to expand. 
This bulging 
increases the length of the field lines, which in turn reduces the 
electric currents that have been induced by the photospheric motions. Analytical 
arguments \citep{Aly85,KmilStu89,Stu95} and numerical simulations 
\citep{MikLin94,Roume94,Amari96a,Amari96b,Aula05a} have shown that, in ideal MHD, 
the expansion-driven current decrease eventually dominates the shear-driven 
current increase. This effect prevents the magnetic field from reaching any loss 
of equilibrium. 
%
%

In all 3D cases, no undriven expansion and therefore no eruption occurs. There are two 
counter-examples only, in 3D \citep{Torok03,Rach09}. But those may not be applicable 
to prominence eruptions. Indeed the related loss of equilibrium there develops 
when the flux rope has strongly expanded, long before the eruption. In 2D, 
shearing motions can easily produce eruptions if reconnection is allowed 
\citep[see][that are further discussed below]{MikLin94,Amari96a,Jaco06}. 
But this behavior has never been reproduced in 3D, except maybe in \citet{Arch08a}. 
All these results suggest that, in general, in 3D, simple line-tied 
shearing/twisting motions alone are not sufficient to drive an eruption. Nevertheless, 
line-tied motions provide a natural process to enhance the departure from non-potentiality 
that is required to power prominence eruptions. 

\subsection{Twisted flux emergence}

Electric current and magnetic pressure can also be directly injected into the 
corona by the emergence through the photosphere of twisted flux ropes that 
rose through the convection zone \citep{Emonet98,JouveBrun09} . 

Some ``kinematic flux emergence'' simulations do achieve this. There the 
emergence is prescribed as time-dependent boundary conditions for the magnetic 
field in a line-tied photospheric boundary, and the whole flux rope can be allowed 
to emerge from the photospheric boundary into the corona. Such simulations indeed 
lead to eruptions 
\citep{FanGibson04,Amari04,Amari05,FanGibson07,Fan10}. 
A clear result came from the careful analysis of some of those. There, eruptions 
have been unambigously shown to be attributed to the torus instability, as 
shown by \citet{FanGibson07} and later by \citet{Fan10}. The former and latter 
constitute the first and third report, respectively, of a simulation that involved 
a torus-unstable flux rope that was gradually formed in the corona, and not prescribed 
as initial conditions as in the first MHD simulations of the torus instability. 

%
%
Unfortunately, simulations of twisted flux emergence through a stratisfied 
medium (hence, non-kinematic emergence) show that, due to the weight of 
photospheric plasma which is trapped in its lower windings 
the flux rope hardly emerges as a whole \citep{Fan01,MagaraLong01,Arch04, Arch09} 
as it does in the kinematic models. Unless the flux rope is not strongly curved 
\citep[e.g. as in][]{MacT09b}, the only way for the lower part of the flux rope 
to emerge is to dispose of the dense plasma trapped in the photospheric dipped 
portions of the field. According to the ``resistive flux emergence model'' this 
may take place through magnetic reconnection 
photospheric \textsf{U}-loops \citep[e.g.][]{Pariat04,Isobe07}. This difficulty 
still raises questions about the results of the kinematic 
simulations. 

Nevertheless, a few non-kinematic simulations of flux emergence 
have successfully produced eruptions, using different codes and initial 
conditions \citep{Manch04,Arch08a,Arch08b,MacT09c,Arch12}. 
But the physical 
mechanism that drive eruptions in these simulations 
remains unclear. Some self-induced shear flows in the photosphere may 
cause eruptions \citep{Manch04}. Magnetic reconnection with an ambient horizontal 
coronal field seems to trigger eruptions \citep[][]{Arch08b,MacT09c}, like in 
the breakout model (see Sect. 4.2). But eruptions are not always 
successful with this process \citep{MacT09a,Leake10}. The development of low-altitude 
magnetic reconnection within the emerging fields could also cause, or at least 
contribute to, the eruption of a newly-formed flux rope \citep{Manch04,Arch08a}, 
like in the tether-cutting model (see Sect. 4.1). Finally, the relative strength 
of the overlying confining arcades as compared to that the emerging rope appears 
determining \citep{Arch12}, maybe like in the axial-flux increase models (see Sect. 
3.1). So more investigation is required in terms of physical analysis. One other 
issue concerns the too small sizes of the modeled flux ropes in these simulations, 
relative to the thickness of the modeled photosphere. 


%
%

\section{MHD models based on decreasing magnetic tension}

Instead of increasing 
the current to a value $I \ge I_c$ 
an alternative approach 
is to 
reduce the restraining tension of coronal 
arcades 
which overlie initially stable current-carrying magnetic 
fields. Most eruption models actually fall into this class. 

\subsection{Tether-cutting}

Magnetic tension can decrease due to the breakdown of ideal MHD in 
the vertical current-sheet that forms within a shearing arcade 
\citep{AmariAly90,ForbesIsen91}, resulting in  magnetic reconnection 
that eventually forms flare loops and ribbons in the wake of the CME, 
i.e. {\em below} the current-carrying field lines. 
This non-ideal effect creates and feeds a twisted envelope around 
the initial current-carrying fields, from the flux of the overlying 
arcades. So the flare reconnection is an efficient process 
for reducing the downward tension of the arcades: it can ``cut the 
tethers'' \citep{Stur89}. 
This process is self-sustaining, since the more the flux rope rises during the 
eruption, and the more reconnection happens, the weaker is the restraining tension, 
so the more the flux rope can rise \citep{Moore92,Moore01,Nishida13}. 
Thus, in principle, it can become explosive. 

The tether-cutting effect alone has indeed 
been shown to trigger and to drive eruptions in 2.5D cartesian \citep{Amari96a} 
and axisymmetric \citep{MikLin94,Jaco06} MHD simulations. 
Early tether-cutting reconnection has also been found to sustain the formation 
of twisted envelopes in 3D MHD simulations of sheared arcades \citep{Devore00}, 
of flux cancellation \citep{Aula10} and of kinematic flux emergence \citep{Fan10}. 
But it did not cause the eruption, when there was one, in any of these 
simulations. Also this reconnection there tends to stall when the 
photospheric driving is supressed during non-eruptive stages. 
Still, the late onset of this reconnection clearly accelerates eruptions in some 
2.5D and 3D MHD simulations. But these eruptions were previously initiated by another 
mechanism, such as the magnetic breakout process 
\citep[][as described hereafter]{Lynch08,Karpen12} and ideal instabilities 
\citep{Nishida13}. 
Some 3D flux emergence simulations did report a qualitative role for 
tether-cutting reconnection in their eruptions \citep{Manch04,Arch08a}. But 
they did not show that it was explicitly driving the eruptions. 

So the tether-cutting has never been proven to initiate, alone, an eruption in 
any 3D simulations. This negative result was found (but rarely 
written) by independent groups using different codes. This raises strong 
doubts about the validity of 
the tether-cutting as an eruption driving mechanism. However, this reconnection 
is obviously an important aspect of every solar eruption. Indeed it provides 
an extra-acceleration to the erupting prominence and, of course, it releases 
a lot of magnetic energy and it produces the most energetic particles in 
the flare that develops in the wake of the CME \citep{Masson13}. 

\subsection{Magnetic breakout}
 
A new idea was proposed by \citet{Ant99}, for lowering the flux and 
the tension of the overlying arcades, by invoking 
magnetic reconnection occurring at a magnetic null point, being  
located at high altitude {\em above} the current-carrying field lines. 

Observationally, this model requires a quadrupolar topology for the 
photospheric magnetic field. This condition can be satisfied in many 
active regions, especially young ones  \citep[see e.g.][]{Ugarte07}. 
But is not guaranteed for older decaying active regions that look 
bipolar \citep[see e.g.][]{vanDriel03}, although large 
remote connections may still be invoked. Theoretically, 
\citet{Ant99} proved for axisymmetric systems that this ``magnetic breakout'' 
alone can drive eruptions, provided that the onset of null point reconnection 
was delayed during the slow energy build-up phase, and that 
the rate of reconnection was slow enough during the fast eruptive 
phase. As for the tether-cutting reconnection, the breakout reconnection 
could create a feedback-loop, leading to an explosive behavior, hence 
to an eruption. In addition, \citet{DeVoreAnt05} found that the efficiency 
of this mechanism also depends on the ratio of the magnetic fluxes located 
above and below the null point: if the flux of the largest overlying 
arcades is too weak (resp. too strong), there is not enough (resp. too 
much) flux to reconnect for the breakout mechanism to be sustained 
long enough for a full eruption. 

Full simulations of the breakout were first achieved in 2.5D MHD simulations 
\citep{MacNeice04}, including with very high spatial resolutions 
\citep{Karpen12,Lynch13} and with the solar wind \citep{vandH07,Masson13}. 
A key difference with the tether-cutting model, 
though, is that the breakout was also found to occur in a 3D line-tied simulation 
\citep{Lynch08}, and very probably in a 3D flux emergence simulation \citep{Arch08b}. 
Also, this original MHD model found unambiguous support in several observational 
analyses \citep[e.g.][]{Aula00,Sterling01,GaryMoore04,Ugarte07}. 

So the breakout mechanism appears as a robust process to initiate and drive solar 
eruptions, although it has several requirements that prevents it from 
being general \citep{DeVoreAnt05}, and it may require a relatively strong flare 
reconnection to produce a fast eruption \citep{Karpen12,Masson13}. 

\subsection{Side-reconnections and remote couplings}

Some other models also explain eruptions through coronal reconnection, 
which occurs aside of the prominence instead of below or above it. In several 
cases this 
reconnection can increase the length of the overlying arcades, and lower 
their tension. 

Eruptions driven by this process were modeled in the context of small-scale 
flux emergence in the vicinity of the flux rope, in the 2D electric-wire paradigm 
\citep{Line01}, in 2.5D MHD simulations \citep{Chen00} and recently in 3D 
simulations \citep{Kusano12,Tori13}. It was also found to operate in 3D MHD 
models of interacting active regions \citep{Jacobs09}, and possibly to trigger 
sympathetic eruptions in models where several current-carrying flux tubes 
are included \citep{TorokP11,Lynch13}, in line with the concept proposed 
by \citet{Sch11b} and further developed by \citet{Sch13}.

The dominant mechanism that drive 
eruptions in these side-reconnection models is still uncertain. \citet{Chen00} attributes 
the eruption to the tether-cutting reconnection triggered by the side-reconnection. 
The broad coverage of the paramater space achieved by \citet{Kusano12} shows 
that the eruptivity strongly depends on the magnetic field configuration. 
\citet{Sch13} and \citet{Lynch13} argue that eruptions are triggered sympathetically 
because the corona is constantly reconfiguring from the previous eruption. 
\citet{Line01} show that an ideal loss of equilibrium is 
triggered in the new system that results from the appearance of a 
new bipole. And finally \citet{TorokP11} show that the coronal reconfiguration that 
results from high-altitude reconnection, actually leads pre-eruptive flux ropes 
that are almost torus-unstable to enter the instability regime and then erupt 
one after the other. 

So, like in flux emergence models, a loss of equilibrium /torus instability can be 
triggered by remote reconnections that result in small-scale or large-scale couplings. 
But other interpretations for the cause of the eruptions have also been proposed. 

\subsection{Converging motions}
 
Models driven by photospheric motions that converge toward polarity inversion 
lines, above which prominences are located, have also been considered.
Quasi-static theory shows that reducing the length-scale of the photospheric 
magnetic field also reduces the magnitude of the coronal field at large heights, 
and makes the field drop faster with height. A priori, both can facilitate the 
torus instability. 

This is strongly suggested by the landmark electric-wire model by 
\citet{ForbesPriest95} and by the MHD simulations of \citet{TorKli07}. They explored 
eruptive behaviors, by making several independent calculations for different 
ratios between the current or the height of the pre-eruptive flux rope, and the 
horizontal extent of the surrounding photospheric bipolar field. 

Eruptions of current-carrying 
fields subject to 
dynamically-treated converging motions, have also 
been found in MHD simulations, 
both in 2.5D \citep{Inhester92} and in 3D \citep{Amari03a}. 
Recently, a direct MHD simulation of an observed event, forced by 
ideal converging motions, also produced an eruption \citep{Zuc12}. 

Such motions are frequently observed at the Sun's surface. This makes 
the model appealing. Some questions remain open, though: very extended 
motions as used in the models are rarely observed; and the physical 
mechanism that actually drives the eruptions has not 
yet been firmly identified in the MHD models. 

\subsection{Decreasing photospheric magnetic field}

This class of models can somehow be viewed as the exact opposite as the 
axial flux increase models. 
They rely on a homogeneous magnetic field decrease in an extended 
section of the photosphere around the flux rope. This decrease is 
imposed, either by reducing the magnetic momentum of the external subphotospheric 
magnetic field sources \citep[see e.g.][]{Lin98}, or by prescribing adequate 
horizontal electric fields in the photosphere \citep[see e.g.][]{Amari00}. 
In axisymmetric geometry, this process produces eruptions of detached flux ropes, 
that are either pre-existing \citep{Lin98}, or slowly formed during the magnetic 
field decrease \citep{Linker03,Ree10}. It can also form and trigger the eruption 
of line-tied flux ropes of various sizes in 3D \citep{Amari00,Line02,Linker03}. 

Qualitatively, the origin of the eruptions can be directly attributed to 
the diminishing of the coronal restraining tension, that naturally results 
from the gradual disappearance of the photospheric magnetic flux. Quantitatively, 
the eruptions occur when the diminishing magnetic energy of the fully open field 
reaches down to a value that is equal to that of the current-carrying fields, as 
identified by \citet{Amari00}. This interpretation is particularly interesting 
because it provides a very clear eruption threshold in the correct MHD paradigm (like 
the torus instability does in the electric-wire paradigm). 

The physical validity of these models is still debated, as 
it is difficult to find a self-consistent MHD process that diminishes 
the photospheric magnetic field over large areas. \citet{Amari00} qualitatively 
noted that flux rope emergence can actually lead to an apparent 
flux decrease on the side rope, after the emergence of the rope axis. But it 
is unclear whether this process produces the magnetic field decrease 
as required for an eruption. \citet{AmariV10} quantitatively calculated 
that photospheric flows that mimic flux dispersal in decaying active regions 
(as described below) can account for the prescribed flux decrease. 
But this interpretation requires flows that accelerate to infinite speeds 
towards the polarity inversion line, which may be problematic. 


\subsection{Flux dispersal and cancellation}

The ``flux-cancellation'' model is based on the observed long-term evolution of 
magnetic flux concentrations in the photosphere, within or between 
bipolar active regions \citep[e.g.][]{WangN89,Dem02,vanDriel03,Schmieder08,Green11}. 
Over time-periods of days to months, depending 
on their sizes, flux concentrations disperse and spread in all directions. 
Their apparent diffusion leads their peak and mean magnetic field magnitude 
to decrease, while their total magnetic flux slowly and weakly decreases through 
local flux convergence and cancellation at polarity inversion lines, right 
below prominences. 

The landmark references for this model are \citet{vanBalle89} and \citet{ForbesIsen91}. 
They showed that converging motions and flux cancellation combined all together (with 
no flux dispersal or decrease) yield the gradual formation of a flux rope through 
a tether-cutting-like photospheric reconnection, that involves coronal arcades 
that have previously been sheared in a 2.5D geometry. On the long run, the rope 
grows in size and in altitude until it erupts, as calculated in 2.5D 
electric-wire models \citep{ForbesIsen91,IsenForbesDem93}. 

By treating the large-scale decay of the photospheric magnetic field with an extra 
photospheric diffusion term in the induction equation \citep[as introduced by][]{WangN89},
3D flux ropes were also found to form and erupt, firstly by \citet{Amari03b} with 
MHD simulations, and later by \citet{MacKayvanBalle06} and \citet{Yeates09} with 
magneto-frictional simulations. These results were also found in non-symmetric MHD 
models \citep{Aula10,Aula12,Pagano13}, and in symmetric models in which the flux dispersal 
was instead treated by line-tied flows diverging from the center each flux concentration 
\citep{Amari11}. 

The magnetic flux decrease model (see Sect. 4.5) is often regarded as a flux 
cancellation model. But both are physically very different. Firstly, in the flux 
cancellation model the magnetic flux decreases locally because of magnetic field 
annihilation at the inversion line \citep{WangN89}. That is different than a flux 
decrease induced by a diminishing magnetic field over a large area. Secondly, the flux 
cancellation model does not involve a dimishing of the open field energy down 
to the magnetic field energy of the pre-eruptive field \citep[see][]{Amari03b}. 

Detailed analysis of one MHD simulation, and its comparison with electric-wire 
models, showed that ``photospheric flux-cancellation and tether-cutting coronal 
reconnection do not trigger CMEs in bipolar magnetic fields, but are key pre-eruptive 
mechanisms for flux ropes to build up and to rise to the critical height above the 
photosphere at which the torus instability causes the eruption'' \citep{Aula10}. 

\section{Discussion}

A large variety of storage-and-release eruption models have been developed 
during the last forty years. At first sight they look similar to each 
other. Indeed, they predict similar observable features, and they share 
common physical ingredients. But they also contain important 
differences, either in their equations, their geometries, and their prescriptions. 
So they have often been opposed to each other. This emulation fostered fine-tuned 
developments and analyses, up to a stage at which they can now be classified 
and compared with one another, so that {\em the physical mechanisms} that 
initiate, drive, and contribute to prominence eruptions may now be identified 
independently of {\em the models} themselves. 

When acknowledging that prominence eruptions occur once the magnetic pressure 
explosively wins over 
the magnetic tension (exerted on the system by the overlying coronal 
arcades), then the present review along with that of \citet{Schmieder13} 
suggest that, to date, no more than two physical mechanisms can initiate 
and sustain this explosive loss of force balance. 

The ideal loss of equilibrium of a flux rope is  
the first mechanism. 3D models have shown that the prominence 
flux rope does not actually need to be very twisted: the mechanism works with 
ropes that have less than one turn. The eruption there occurs once the rope axis 
has reached an altitude, above which all stable equilibria cease to exist. 
The threshold is reached when the magnitude of magnetic fields of the overlying 
arcades
decrease 
faster with height than the magnetic pressure which pushes the flux rope 
upwards, which also decreases with time during the rise of the flux 
rope. The process was first proposed by \citet{vantend78}, and it was 
shown by \citet{DemAula10} to correspond to the ``torus instability'' 
first proposed by \citet{Bateman78} in tokamaks, and first revisited 
for solar eruptions by \citet{KliTor06}. 

The removal of the arcades that overlay and confine the prominence 
by means of high-altitude magnetic reconnection is the second mechanism. 
It has initially been proposed to occur at null points, but it may also 
operate at separators and quasi-separatrix layers. Once the reconnection 
has begun, it transfers overlying arcades into connectivity domains 
that are located aside of the prominence. So the amount of magnetic flux 
that overlays the prominence is reduced. The associated diminishing of 
the confinement makes the prominence rise to larger altitudes. This 
provides a loop-feedback on the high-altitude reconnection, so that 
an eruption can occur. This process, called the ``magnetic breakout'', 
was first proposed by \citet{Ant99}. The efficiency of the breakout 
requires the magnetic fluxes located above and below the reconnection 
region be comparable in magnitude \citep{DeVoreAnt05}.

To date, only the torus instability and the magnetic breakout were found to 
occur in many different 3D MHD simulations. 
The torus instability has been identified to cause eruptions with prescribed 
unstable flux ropes \citep{Roussev03,To05,TorKli07,SchrijverE08,Tor10}, 
with kinematic flux emergence \citep{FanGibson07,Fan10}, with flux cancellation 
\citep{Aula10,Aula12}, with sympathetically erupting flux ropes \citep{TorokP11}, 
and with non-linear force-free relaxations \citep{Kliem13,Jiang13}. 
The magnetic breakout mechanism has been shown to operate with shearing bipoles 
in multipolar geometry \citep{Lynch08}, with twisted flux tubes emerging 
through a stratified medium into a pre-existing horizontal field 
\citep{Arch08b,MacT09c}, and with sympathetically erupting sheared loops 
\citep{Lynch13}. So it can be conjectured that eruptions can only be initiated 
and driven by one of these two mechanisms, or their combination. 

It follows that, depending on the solar conditions, all the other processes may  
be considered as different ways to either bring the system to 
the threshold of one of these two mechanisms, or to help making the resulting 
eruption faster. For example, flux emergence 
\citep{Fan10} or flux cancellation \citep{Aula10} can initiate a torus instability. 
Also, reconnection-driven long-range couplings around flux ropes \citep{TorokP11} 
or sheared arcades \citep{Lynch13} can initiate sympathetic torus instabilities 
and magnetic breakouts. And flare reconnection can accelerate eruptions initiated 
by a torus instability \citep{Nishida13} and a magnetic breakout \citep{Karpen12,Masson13}. 


In this line each and every solar process that can contribute to solar eruptions 
should be taken into account, all together with the few physical mechanisms that 
initiate and drive eruptions, so as to reach a comprehensive 
understanding of observed events, and so as to predict the occurrence of future 
events.


\end{document}